\documentclass[a4paper,10pt,twoside]{cpc-hepnp}

\usepackage{multicol}
\usepackage{graphicx}
\usepackage{booktabs}
\usepackage{amssymb,bm,mathrsfs,bbm,amscd}
\usepackage[tbtags]{amsmath}
\usepackage{lastpage}

\newcommand{\nc}{\newcommand}
\nc{\be}{\begin{equation}}
\nc{\ee}{\end{equation}}
\nc{\ra}{\rightarrow}
\nc{\pa}{\partial}
\nc{\parsym} {\stackrel{\leftrightarrow}{\pa}}
\nc{\omg}{\omega}
\begin{document}

\fancyhead[co]{\footnotesize M. Benayoun~ et al: 
Can VMD improve the estimate of the muon $g-2$~? }

\footnotetext[0]{Received XX December 2009}

\title{ Can VMD improve the estimate of the muon $g-2$~? }

\author{%
      M. Benayoun$^1$\email{benayoun@in2p3.fr}%
\quad L. Del~Buono$^1$
\quad B. P. David$^1$
\quad O. Leitner$^1$
}
\maketitle

\address{%
1~(LPNHE des Universit\'es Paris VI et VII, IN2P3--CNRS,  Paris,  75005,  France)\\
}

\begin{abstract}
We show that a VMD based theoretical input allows for a significantly  improved accuracy for the
hadronic vacuum polarization of the photon which contributes to the theoretical estimate
of the muon $g-2$. We also show that the only experimental piece of information in the
$\tau$ decay which cannot be accounted for is the accepted value for
 ${\rm Br}(\tau \ra \pi \pi \nu_\tau)$, while the spectum lineshape is 
 in agreement with expectations from $e^+ e^-$  annihilations.
\end{abstract}



\begin{multicols}{2}

\section{Introduction}

The Hidden Local  Symmetry (HLS) Model\cite{HLS1,HLS2} implements the 
Vector Meson Dominance asssumption within the framework of Effective Lagrangians.
The non--anomalous sector of this model covers annihilation channels 
like $e^+ e^- \ra \pi^+ \pi^-$ or $e^+ e^- \ra K \overline{K} $ and some important
decay channels like $\tau \ra \pi \pi \nu_\tau$. The non--anomalous sector can be 
supplemented with an anomalous sector \cite{FKTUY,WZ,Witten}, allowing\footnote{
From now on $P$ and $V$ denote any pseudoscalar and any vector fields from the
basic SU(3) nonets.} for $\gamma \gamma P$, $\gamma P V$, $P V V$, $\gamma P P P$ and
$V P P P$ couplings. Therefore, annihilation processes like   $e^+ e^- \ra (\pi^0/\eta)  \gamma$,
or  $e^+ e^- \ra \pi^0 \pi^+ \pi^-$ can enter the HLS framework as well as all radiative 
decay  processes  of the form $V \ra P \gamma $  or $P \ra \gamma \gamma$  
 or  also processees like $\eta/\eta^\prime \ra \pi^+ \pi^- \gamma$.

Therefore, the HLS model provides a unified framework valid in the low energy regime up
to the $\phi$ mass region. It encompasses most annihilation  
and decay processes.

However, in order to be confronted with experimental data, the HLS model should be 
equiped with symmetry breaking mechanisms.
  Implementing SU(3) breaking is done using a variant\cite{heath}   
of the BKY mechanism\cite{BKY} in the non--anomalous sector. Breaking of the (nonet)
U(3) symmetry for pseudoscalar  mesons is also an important issue~; it is generated 
\cite{WZWChPT} by determinant term  Lagrangian pieces\cite{tHooft}.  
 The SU(3) breaking of the anomalous Lagrangian 
is done following the scheme proposed by \cite{BGP,BGP2} supplemented 
with a vector field renormalization recently  justified \cite{taupaper}.
This full SU(3)/U(3) breaking of the HLS model, recalled in \cite{taupaper},
has allowed a successfull description of all light meson radiative decays \cite{rad,box}.

A consistent  treatment of the $e^+ e^- \ra \pi^+ \pi^-$  annihilation
and the $\tau \ra \pi \pi \nu_\tau$ decay requires an appropriate
mechanism for Isospin Symmetry breaking (ISB). This has been  defined in
\cite{taupaper} and has improved the description of all processes listed 
above (annihilation and decay processes) as shown in \cite{ExtMod1,ExtMod2}.

\section{How  can VMD improve estimates of $g-2$~?}

Therefore, the HLS model provides a framework able to describe in a unified way an important number
of cross sections\footnote{ We will use 
$e^+ e^- \ra \pi^+ \pi^-$, $e^+ e^- \ra \pi^0 \gamma$, 
$e^+ e^- \ra \eta \gamma$  and  $e^+ e^- \ra \pi^0 \pi^+ \pi^-$.
Instead, the $e^+ e^- \ra K \overline{K} $ cross section will be left aside because
of a still misunderstood problem concerning the ratio of two kaon decay modes of the $\phi$ meson\cite{BGPter}.
} with an additional set of radiative decay modes.
 These play the major role of constraints
in order to determine numerically the  parameters of the SU(3)/U(3)/SU(2) breaking scheme.
The decay $\tau \ra \pi \pi \nu_\tau$ is nothing but an additional
constraint, also subject to ISB effects  usually split
up into short range  \cite{Marciano} and long range \cite{Cirigliano2,Cirigliano3}
 (resp. $S_{EW}$  and $G_{EM}(s)$) corrections. These are only overall rescaling factors.

Within this unified model \cite{ExtMod1}, all relevant data (already listed) depend on 
a very few basic parameters, namely the CKM matrix element $V_{ud}$,
 the electric charge $e$, the pion decay constant $f_\pi$, 
the universal vector coupling $g$, the weak interaction coupling $g_2$ 
and a parameter named $a$, specific of the HLS model \cite{HLS2,heath},  
expected close to 2.  $V_{ud}$, $f_\pi$, and $g_2$  (related
to the Fermi constant $g_2=2 m_W\sqrt{\sqrt{2} G_F}$) are accurately known. 
Therefore, the only parameters to be fitted from data are
 $a$ and $g$. The anomalous 
sector introduces 4 more parameters (named $c_i$ in  \cite{HLS2}) 
in such a way that only two parameters should be determined by fit \cite{ExtMod1}~:
the combination $c_1-c_2$ and $c_3$. 

The U(3)/SU(3) breaking procedure introduces 4 breaking parameters  
determined by only the radiative decays
\cite{taupaper,ExtMod1}~: $z_A$, $z_V$, $z_T$ and $x$. Some of these have a clear
physical
meaning. Indeed,  $z_A =[f_K/f_\pi]^2$ is the squared ratio of the kaon and pion 
decay constants.  $x$ is the nonet symmetry breaking parameter,  tightly
related with the pseudoscalar mixing angle  in the octet--singlet basis \cite{box,taupaper}
$\theta_{PS} \simeq -10^\circ$. More important for the present purpose is the ISB
breaking scheme which introduces more parameters \cite{taupaper,ExtMod1} to
be fitted  and is sketched below.

Our extended model \cite{ExtMod1} can provide a global fit to the whole set of data
listed above. Stated otherwise, the parameters given above underly a physics
content common to a very large number of annihilation and decay channels. Therefore,
our overconstrained parametrization of the VMD physics allows for a global 
overconstrained fit. Then, if these constraints are well accepted by the
data, parameter values and the parameter error covariance matrix will be defined
with high accuracy.  This should reflect in better estimates of the
various contributions of the photon hadronic vacuum polarization (HVP)  to $a_\mu$.
For this purpose, one only relies on the description quality of the annihilation cross sections
and on the consistency of the various data sets with each other.

The quality of the description of the various cross sections gives also a hint on the
quality of the estimates these allow for  $a_\mu$. As stated above, the limit
of validity of the HLS model extends to slightly above the $\phi$ mass. 
However, this $s$ region contributes more than 80\% to
the numerical value for  $a_\mu$ and the corresponding uncertainty
 is as large as $\simeq 35\%$  of the total  $a_\mu$ uncertainty.
Therefore, even if limited, the expected improvements may have important
consequences concerning the physics of $g-2$.

\section{Breaking of the isospin symmetry: Vector field mixing}

Concerning the sector of neutral vector mesons, at leading (tree) order
the ideal fields $\rho^0_I$, $\omg_I$ and $\phi_I$ which enter the 
HLS Lagrangian -- as any VMD Lagrangian -- are mass eigenstates with
resp. masses $m_\rho^2=m_\omg^2=m^2$ and $m_\phi^2=z_V m^2$ 
($m^2=a g^2 f_\pi^2$). However, at one loop order, the Lagrangian piece
\begin{eqnarray}
\label{eq1}
{\cal L}_1  &=& \dfrac{ia g}{4 z_A} \left \{ 
\left [\rho^0_I+\omg_I-\sqrt{2} z_V \phi_I \right ] 
K^- \parsym K^+ \right.  \nonumber\\[1mm]
&& + 
\left.
\left [\rho^0_I-\omg_I+\sqrt{2} z_V \phi_I \right ] 
K^0 \parsym \overline{K}^0 
\right\}
\end{eqnarray}
induces transitions among the ideal 
vector meson fields $\rho^0_I$, $\omg_I$ and $\phi_I$ through kaon loops\footnote{
Actually, in the more complete Lagrangian, $K^* \overline{K}^*$
and $K K^*$ loops come in  complementing the kaon loops along the same lines
\cite{taupaper}.}. 
Therefore, at one loop order, the ideal fields are no longer mass eigenstates
and thus do not coincide any longer with the physical
$\rho^0$, $\omg$ and $\phi$ fields which, instead, must be mass eigenstates.  At one loop
order, the squared mass matrix $M^2$ for the field triplet ($\rho^0_I$, $\omg_I$, $\phi_I$)
is given by Eq. (12) in \cite{taupaper} and its eigensystem can be constructed perturbatively. One can
define  3 mixing functions \cite{ExtMod1} ~:  
$\alpha(s)$,  $\beta(s)$, $\gamma(s)$ which can be 
considered as complex "angles" and are function of  $s$, the squared momentum flowing through 
the vector meson 
line. 
$\alpha(s)$, $\beta(s)$ and   $\gamma(s)$ describe resp. the $\rho^0-\omg$,
$\rho^0-\phi$ and $\omg-\phi$ mixings. These angles \cite{ExtMod1}, functions
of the kaon loops and of the pion loop, contain subtraction polynomials to be fitted
using experimental data. The relationship between ideal and physical fields can be
written in terms of these angles~:
\be
\left (
\begin{array}{lll}
\rho_I^0\\[0.1cm]
\omg_I\\[0.1cm]
\phi_I
\end{array}
\right ) = 
\left (
\begin{array}{cll}
\displaystyle  \rho^0 -\alpha(s) ~\omg+\beta(s) ~\phi  \\[0.1cm]
\displaystyle  \omg +\alpha(s) ~ \rho^0+\gamma(s)  ~\phi \\[0.1cm]
\displaystyle \phi -\beta (s) ~ \rho^0 -\gamma(s)  ~\omg
\end{array}
\right ) 
\label{eq2}
\ee
Therefore, the vector meson mixing induced by loop corrections 
being $s$--dependent, is a quite important feature.
This transformation propagates to the interaction terms. For instance,
the term describing the interaction of a pion pair with vector mesons
becomes~:
\begin{eqnarray}
\label{eq3}
\dfrac{ia g}{2} \rho^0_I \pi^- \parsym \pi^+
\Rightarrow \dfrac{ia g}{2} \left[
 \rho^0-\alpha(s)  ~\omg +\beta(s)  ~\phi
\right]
\end{eqnarray}
clearly exhibiting the origin of the isospin 1 part of the physically
observed $\omg$ and $\phi$ fields.

The $\gamma-V$ transition term is also interesting. It becomes~:
\begin{eqnarray}
\label{eq4}
-e a g f_\pi^2 \left [
\rho_I^0 +\dfrac{1}{3} \omg_I-\dfrac{\sqrt{2}z_V}{3} \phi_I
\right ]\cdot A \Longrightarrow \nonumber\\[1mm]
-e  \left [
f^\gamma_\rho(s)~\rho^0 +f^\gamma_\omg(s) ~\omg - f^\gamma_\phi(s)\dfrac{\sqrt{2}z_V}{3} ~\phi
\right ]\cdot A
\end{eqnarray}
where $f_V^\gamma(s)= agf_\pi^2[1+{\cal O}(\alpha(s), \beta(s),\gamma(s))]$
has well defined correction terms \cite{taupaper,ExtMod1}. The electromagnetic field
is denoted by $A$. 

The most interesting feature
 here
concerns the $\rho$ meson which then gets different transition amplitudes
 to the $\gamma$ and $W$ fields, One can, indeed, show that the amplitude ratio
 is~:
 \begin{eqnarray}
\label{eq5}
\dfrac{f^\gamma_\rho}{f^W_\rho}= \left [
1 + \dfrac{\alpha(s)}{3}+\dfrac{\sqrt{2} z_V}{3} \beta(s)\right ]~~~~~,
~~(f^W_\rho=a g f_\pi^2)
\end{eqnarray}
where the $s$--dependent terms represent  the isospin 0 part of
the $\rho^0$ meson inherited from its $\omg_I$ and $\phi_I$ components. 
This makes different the interaction of the $\rho^0$ and $\rho^\pm$
fields with resp. the $\gamma$ and $W$ gauge fields. 

Therefore, our isospin breaking scheme results in physical vector fields
which are mixtures  of definite isospin components and their exact content
is $s$--dependent.

\section{Sketching the global fit to data}
\label{globalFit}

The  cross sections for $e^+ e^- \ra \pi^+ \pi^-$, 
$e^+ e^- \ra \pi^0 \gamma$,
$e^+ e^- \ra \eta \gamma$  and  $e^+ e^- \ra \pi^0 \pi^+ \pi^-$
have been worked out in \cite{ExtMod1} together with the expressions
for the relevant set of decay partial widths. The expression for the
$\tau \ra \pi \pi \nu_\tau$ spectrum has been computed in \cite{taupaper}
and can also be found in \cite{ExtMod2}. The corresponding formulae have been 
implemented within a computer code aiming at performing a (simultaneous) global
fit to all existing relevant data.

All existing $e^+e^-$ annihilation data samples have been considered in the 
context of our global fit method. 
For the $\pi^+ \pi^-$ final state, this covers 
the former data sets collected in \cite{Barkov} and in \cite{DM1} and the more
recent ones collected at Novosibirsk
\cite{CMD2-1995,CMD2-1995corr,CMD2-1998-1,CMD2-1998-2,SND-1998}.  
All existing data sets with the $(\pi^0/\eta) \gamma$ final states have also
been considered  \cite{CMD2Pg1999,CMD2Pg2005,SNDPg2000,SNDPg2003,SNDPg2007}.

For the $\pi^0 \pi^+ \pi^-$ annihilation channel, the main available 
data sets have been  provided by CMD--2
\cite{CMD2-1995,CMD2-3p-1995,CMD2-3p-1995,CMD2-3p-1998,CMD2-3p-2006}
and SND \cite{SND3pionLow,SND3pionHigh}.
These have been considered along with
older data sets \cite{ND3pion-1991,CMD3pion-1989}~; only the very old
data set from \cite{DM13pion-1979} has been eliminated because it was not 
clear how to account precisely for its systematics. 

Actually, after analyzing the scale uncertainties
claimed for the  CMD--2 and SND three pion data sets, we were led to leave aside 
\cite{ExtMod1} also the SND data sets \cite{SND3pionLow,SND3pionHigh}.  

Finally, the $\pi^+ \pi^-$  KLOE data set,
collected at Frascati using the ISR method and reanalyzed recently
\cite{KLOE-2008}, has been included in the data sets considered.

Concerning the $\tau \ra \pi \pi \nu_\tau$ spectra,  
 we considered those from CLEO  \cite{Cleo}, 
ALEPH \cite{Aleph} and BELLE \cite{Belle}. 
These data sets will be commented
with some details in Section 6.

Full information about the fit properties and  qualities can be found in
\cite{ExtMod1,ExtMod2} and are not presented here because of lack of place.
Let us only mention that they are always affected by very good probabilities.

\section{Improved estimate of the photon HVP}
\label{gMoins2}

In order to estimate the various contributions of the photon 
 HVP to $a_\mu$ for $s\leq 1$ GeV,  
we followed a specific procedure~:
\begin{itemize}
\item Use always all  the $e^+e^-$ annihilation data samples essentially collected
at Novosibirsk 
\cite{Barkov,DM1,CMD2-1995,CMD2-1995corr,CMD2-1998-1,CMD2-1998-2,SND-1998},
\cite{CMD2Pg1999,CMD2Pg2005,SNDPg2000,SNDPg2003,SNDPg2007}, 
\cite{CMD2-1995,CMD2-3p-1995,CMD2-3p-1995,CMD2-3p-1998,CMD2-3p-2006,ND3pion-1991,CMD3pion-1989} 

\item Use always the various partial widths of types $VP\gamma$ and   $P \ra \gamma \gamma$
as reported in the Review of Particle Properties \cite{RPP2008}. These play a
crucial  role in order to overconstrain our model parameter values.

\item Examine the effect of the $\pi^+ \pi^-$ KLOE data \cite{KLOE-2008} separately, because 
the fit properties of this sample are not fully satisfactory.

\item Add as further constraints, separately or altogether, the $\tau$ data
from  C (CLEO   \cite{Cleo}), B (BELLE \cite{Belle}) and/or A (ALEPH \cite{Aleph}),
in order to exhibit the  specific influence of each.
\end{itemize}

In the comparison with experimental data, we  focus in the following
on the contribution of the pion loop only  ({\it i.e.}  $a_\mu(\pi \pi)$), integrated
between $\sqrt{s}=0.630$ GeV and $\sqrt{s}=0.958$ GeV. Indeed, most
experimental groups have published their estimates for  $a_\mu(\pi \pi)$
in this reference energy range. As these experimental results are corrected
for final state radiation (FSR) effects, we do alike.

In order to check our method and illustrate its effect, 
we have first run our code using each of the data
sets from \cite{CMD2-1995corr}, \cite{CMD2-1998-1} and \cite{SND-1998} in isolation,
{\it together with our full set of radiative decay information} (17 pieces).
The results derived from the fitted pion form factor are reported in the first 3 lines of Table \ref{T1}
and the errors shown  are the total uncertainties. Indeed, the fit is done with a procedure
combining appropriately \cite{ExtMod1,ExtMod2} statistical and systematic errors. 

\begin{center}
\tabcaption{ \label{T1}  Our estimates for $10^{10}~a_\mu(\pi \pi)$
and the corresponding experimental values from resp. \cite{CMD2-1995corr}, \cite{CMD2-1998-1},
\cite{SND-1998}. The last two lines are averages proposed by \cite{Davier}.
}
\footnotesize
\begin{tabular*}{80mm}{c@{\extracolsep{\fill}}ccc}
\toprule Data Set & Exp. Value   & Reco. Value  & Prob. \\
\hline
CMD2 (1995) \hphantom{0}  & $362.1 \pm 2.4 \pm 2.2$ & \hphantom{0}$362.9\hphantom{0}^{+3.1}_{-4.5}$ & 51\% \\
CMD2 (1998) \hphantom{0}  & $361.5 \pm 1.7 \pm 2.9$  & \hphantom{0}$362.2\pm2.1$ & 49\% \\
SND (1998) \hphantom{0}  & $361.0 \pm 1.2 \pm 4.7$  & \hphantom{0}$361.0 \pm 2.1$& 99\% \\
\hline
NSK (all) \hphantom{0} &  $360.2 \pm 3.0_{tot}$\hphantom{00} & \hphantom{0}$361.7 \pm 1.3$ & 48\%\\
\hline
\hline
NSK +KLOE \hphantom{0} &  $358.5 \pm 2.4_{tot}$\hphantom{00} & \hphantom{0}$362.1\pm 1.1$ & --\\
\bottomrule
\end{tabular*}
\end{center}

One clearly observes an important improvement of the accuracy
following from having built, for the first time,
a working model which simultaneously fits the radiative decays and the annihilation data. 
Comparing the results obtained using each of the CMD--2 and SND data sets
in isolation and altogether, the effect expected from an increased statistics is 
observed with its expected magnitude. The net effect is a factor of $\simeq 2$
improvement of the uncertainty.
As will be seen shortly, this is also due to the fact that the uncertainties 
(and biases) within the data sets just quoted are well under control.
The  last data column gives the probability of the underlying fit to the
pion form factor and the decay data. The fit probability
of the SND data clearly reflects a too conservative estimate of their systematics.

KLOE data \cite{KLOE-2008} help in slightly improving estimates
 at the expense, however,
of a poor fit probability, essentially due to a (still) poor control of the systematics 
within this data set.

Our favorite estimate of $a_\mu(\pi \pi)$ (fourth line in Table \ref{T1}) compares
favorably to the newly issued experimental results produced from recent pion form 
factor data collected
using the ISR method by the KLOE Collaboration \cite{KLOE-2009} ($10^{10}~a_\mu(\pi \pi)= 356.7 \pm 0.4 \pm 3.1$)
and by the BaBar Collaboration \cite{Davier2} ($10^{10}~a_\mu(\pi \pi)= 365.2 \pm 2.7$). 
These two new measurements illustrate that one needs motivated theoretical input
in order to take a full profit of the new high statistics data sets. Indeed,  
\cite{Davier2} proposes an average of the four experimental values given in 
Table \ref{T1} (\cite{CMD2-1995corr,CMD2-1998-1,SND-1998,KLOE-2008}) and of the BaBar estimate 
\cite{Davier2} over the same energy range~; using a sophisticated statistical method, they
get $10^{10}~a_\mu(\pi \pi)= 360.8 \pm 2.0_{tot}$.

Comparing this average with our fit value (last line in Table \ref{T1}) -- which does
not use the (not yet public) BaBar data -- is interesting. Indeed, it shows that the increased
statistics  provided by the ISR method at DAPHNE and BaBar has not  allowed a real breakthrough
in the accuracy of $a_\mu(\pi \pi)$, because of the systematics  specific to each experiment
and of the difficulties encountered while merging the different data samples. 

Instead, what is illustrated by Table \ref{T1} is that an adequate theoretical input -- like
VMD -- may allow sizable improvements. Of course, the relevance of this input should be (and
actually is) reflected by the global fit qualities \cite{ExtMod1,ExtMod2}.

We do not discuss here the effects of introducing the $(\pi^0/\eta) \gamma$ and
$\pi^+ \pi^- \pi^0$ data~; this has been analyzed in full details in \cite{ExtMod2}.
Let us, nevertheless, mention that these data sets, with poorly known sytematics,
allow to confirm the central values for $a_\mu(\pi \pi)$ without a visible 
improvement of its uncertainty. 

\section{Adding the $\tau$ spectra to the fitted data samples}

As mentioned in Section 4, we only deal with the CLEO (C) \cite{Cleo},
ALEPH (A) \cite{Aleph} and BELLE (B) \cite{Belle} data sets.  
The (C) data set provided by CLEO is actually the normalized spectrum 
$1/N dN/ds$. The absolute normalization  for  $d\Gamma(\tau \ra \pi \pi \nu_\tau)/ds$ 
is determined by a multiplicative factor\footnote{See, for instance, Eq. (7) in \cite{Belle} }, 
the branching ratio ${\rm Br}(\tau \ra \pi \pi \nu_\tau)$. Therefore, the CLEO spectrum
we use is not sensitive to this branching ratio.
As, following the BELLE
Collaboration \cite{Belle}, we allow for a rescaling of the B data set, we
are only marginally sensitive to ${\rm Br}(\tau \ra \pi \pi \nu_\tau)$. Instead, as there
is no reported uncertainty on the normalization of the ALEPH (A) $|F_\pi(s)|^2$
spectrum, we have not allowed any  rescaling for the A data set.

This way to proceed with B and C is not the usual one. Indeed, usually, the  B and C $|F_\pi(s)|^2$
spectra are constructed as their reported normalized spectrum $1/N dN/ds$ multiplied
by the world average value\footnote{As can be concluded from Figure 6 in \cite{Davier}, 
the world average
value for ${\rm Br}(\tau \ra \pi \pi \nu_\tau)$ differs only marginally from the corresponding
ALEPH \cite{Aleph} measurement. } for ${\rm Br}(\tau \ra \pi \pi \nu_\tau)$ \cite{Belle,Davier}.

In the (global) HLS model, the $\tau$ spectrum is determined essentially by the Higgs--Kibble 
$\rho^\pm$ mass
(occuring in the Lagrangian) and by the $\rho^\pm$ coupling to a pion pair. Naming
the $\rho^0$ mass squared $m^2(=ag^2f_\pi^2)$ and $g$ its coupling constant 
to a pion pair, we have defined the corresponding quantities for the $\rho^\pm$
meson by $m^2+\delta m^2$ and $g+ \delta g$. Interestingly, the absolute
magnitude of the $\tau$ spectrum  and the $\rho^\pm$ width are both determined by 
the $\rho^\pm \pi \pi$ coupling constant and then by $g+ \delta g$.
Isospin symmetry breaking effects  specific of the $\tau$ decay 
modify this picture by introducing short range  \cite{Marciano} ($S_{EW}$) and long
range  \cite{Cirigliano2} ($G_{EM}(s)$) corrections which both factor out
and, therefore, contribute to the absolute magnitude of the $\tau$ spectrum.

 \begin{center}
\includegraphics[width=8cm]{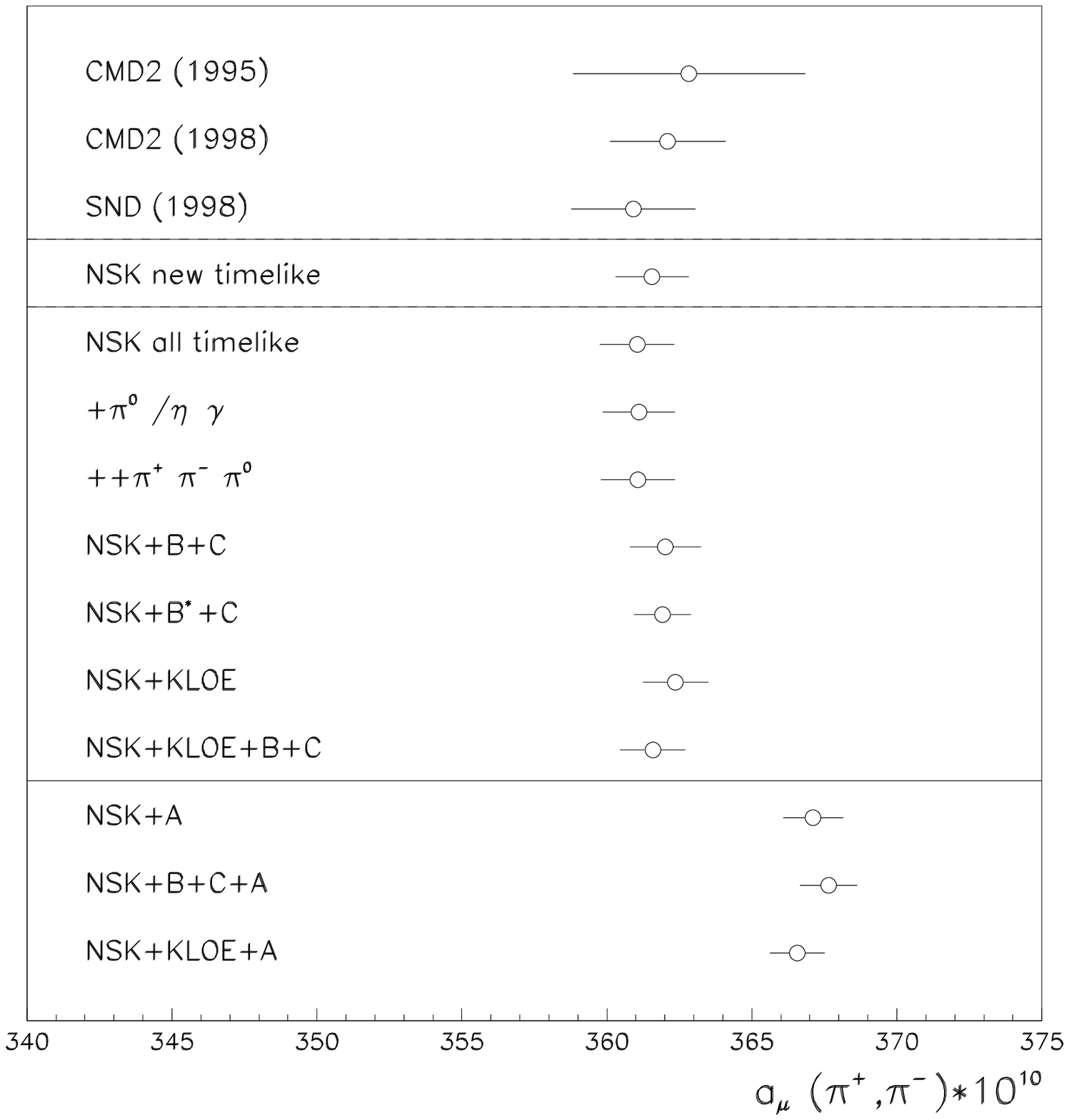}
\figcaption{\label{fig0}   Figure 1. The $\pi \pi$ loop contribution to 
$g-2$ integrated between 0.630 and 0.958 GeV  using our global fit
running with various combinations of data sets. }
\end{center}

We  have first performed fits with 
$e^+e^-$ and $\tau$ data in order to determine $\delta m^2$ and $\delta g$.
It happens \cite{ExtMod1,ExtMod2} that the fits return  $\delta m^2$ and $\delta g$
consistent with zero at a $\simeq 1~ \sigma$ level. Therefore, we do not find significant 
differences between the  $\rho^0$ and $\rho^\pm$ (Lagrangian) masses and couplings.
It thus follows that the difference between the pion form factor in $\tau$ decays
($F_\pi^\tau(s)$)  and the I=1 part of the pion form factor in $e^+ e^-$ annihilations 
($F_\pi^e(s)$) is fully carried by the factor $\sqrt{S_{EW}G_{EM}(s)}$,
which affects the $\tau$ dipion spectrum.

Then, fixing  $\delta m^2=\delta g=0$, we have redone our final fits 
allowing for a rescaling of the B data sample, by varying the set of
data sets (listed in Section 4) submitted to the global fit.

\section{A localized failure of CVC~?}

Our global fits are always fairly good \cite{ExtMod1,ExtMod2}
and result in an overall  rescaling factor for the B data sample 
$1+\lambda$ with $\lambda=(-4.84^{+1.37}_{-0.92}) \%$, in good
correspondence with the BELLE fit result \cite{Belle} which can be
written $\lambda_{BELLE}=-(2 \pm 1 \pm 4)\%$. In this approach, the
C and B data samples are always well described~; the ALEPH spectrum
is reasonably well described below 1 GeV, however more poorly than
the C and B data samples \cite{ExtMod2}. This is partly due\footnote{
Indeed, if the CLEO and BELLE pion form factors are in fairly good 
agreement with each other, they sensitively differ from the ALEPH
form factor in the very low and in the high energy regions, as can be seen
from Figure 12 in \cite{Belle}.} to the fixed absolute normalization
of ALEPH data, {\it i.e.} to the accepted value for 
${\rm Br}(\tau \ra \pi \pi \nu_\tau)$.

At this step, one should note that the HLS model we use, equiped
with symmetry breaking schemes accounts fairly well for~:
\begin{itemize}
\item all $e^+ e^-$ annihilation cross sections listed in Section 4,
\item all partial width decays of the form $P\gamma \gamma$,
$VP\gamma$ and $\eta/\eta^\prime \ra \pi^+ \pi^- \gamma$,
\item the {\it lineshape} of the dipion spectrum in $\tau$ decay, especially
those provided by CLEO and BELLE which are quite similar.
\end{itemize}
Stated otherwise, the single piece of information which does not fit
within this overall picture
is the accepted absolute normalization of the $\tau$ dipion spectrum,
{\it i.e.} ${\rm Br}(\tau \ra \pi \pi \nu_\tau)$. 

If one excludes an experimental bias, one thus needs a specific 
additional breaking effect affecting solely the $\tau$ decay.
However, as this missing piece resembles a global rescaling of the
$\tau$ spectrum, a possible candidate could be a revised $S_{EW}$ factor
numerical value, if relevant\footnote{It is generally assumed that the value found for
$\tau \ra \pi \nu$ \cite{Marciano} coincides with the corresponding
factor for $\tau \ra \pi \pi \nu$. This crucial assumption does
not seem to have been proved.}.

\section{Influence of the $\tau$ spectrum}

Figure \ref{fig0} displays the value for $a_\mu(\pi \pi)$ integrated
along the canonical interval around the $\rho$ peak, as coming from our
(global) fits. The 4 upmost data points are the values shown in Table
\ref{T1}. Thus, the fourth line gives the result derived from
a combined fit to the data given in \cite{CMD2-1995corr,CMD2-1998-1,CMD2-1998-2,SND-1998}. 
The fifth line displays the result coming from the 
combined fit to the $\pi^+ \pi^-$ data sets
just quoted and to the older $\pi^+ \pi^-$ data sets given in \cite{Barkov,DM1}. For the line  indicated
by +$\pi^0/\eta \gamma$ , we have added the corresponding data sets to all $\pi^+ \pi^-$ data.
In order to get the result indicated at the line flagged by ++$\pi^+ \pi^-\pi^0$, the 
corresponding data samples have been considered together with all the previous ones. Concerning 
the  rest of Figure \ref{fig0}, NSK denotes all  $e^+ e^-$ annilhilation data combined 
with KLOE, ALEPH, BELLE, CLEO in the way indicated at the corresponding line.

One can conclude from Figure \ref{fig0}, that all data set combinations submitted to fit
and built up from all $e^+ e^-$  annihilation samples and from the B and C sets
provide quite consistent results. Instead, as shown by the 3 downmost $a_\mu(\pi \pi)$ values,
including the ALEPH data set always provides a shift upwards by 
$ \simeq 5~10^{-10}$. This is almost certainly related with the branching ratio issue discussed 
in Section 7.

\section{Conclusions}

We have proved that a theoretical VMD input permits to significantly improve
the accuracy of predicted value for the muon  $g-2$ value, as clear from Table \ref{T1}.
Some further improvement is reached by adding the $\tau$ spectra, however marginal.
Our VMD input certainly increases the
disagreement between the expected value for the muon $g-2$ and its direct BNL measurement.

Another important remark is that the $\rho$ meson {\it lineshape} observed
in the $\tau$ dipion spectra in perfect agreement with expectations from VMD. The single surviving
issue in our data set, the largest one ever analyzed within  a single model, is solely the value for 
${\rm Br}(\tau \ra \pi \pi \nu_\tau)$, expected slighly smaller than its presently 
accepted value. If not an experimental bias, this may indicate that symmetry breaking
effects in $\tau$ decays are still to be revisited.
Until this issue  is clarified, one 
should consider cautiously the predictions for the muon
$g-2$ provided by the $\tau$ dipion spectrum, especially those depending on the
absolute scale of this spectrum.  

\end{multicols}

\vspace{-2mm}
\centerline{\rule{80mm}{0.1pt}}
\vspace{2mm}

\begin{multicols}{2}

\end{multicols}

\clearpage


\begin{thebibliography}{90}

\vspace{3mm}

\bibitem{HLS1} Bando  M,  Kugo T, Yamawaki  K, Phys. Rept., 1988, {\bf 164}: 217--314

\bibitem{HLS2} Harada M, Yamawaki T, Phys. Rept., 2003, {\bf 381}: 1--233

\bibitem{FKTUY} Fujiwara  T, Kugo T, Terao  H, Uehara  S, Yamawaki  K, 
Prog. Theor. Phys., 1985,  {\bf 73}:926--941

\bibitem{WZ} Wess J, Zumino B, Phys. Lett., 1971, {\bf 37}: 95--97

\bibitem{Witten} Witten  E, Nucl. Phys., 1983, {\bf B223}:422--432
 
\bibitem{heath} Benayoun  M,  O'Connell H  B, Phys. Rev.,  1999, {\bf D58}: 074006

\bibitem{BKY} Bando  M,  Kugo T, Yamawaki  K, Nucl. Phys., 1985, {\bf B259}:493--502

\bibitem{WZWChPT} Benayoun  M, DelBuono, L, ,O'Connell, H  B, Eur. Phys. J.,
2000, {\bf C17}:593-610

\bibitem{tHooft} 't Hooft G, Phys. Rept., 1986, {\bf 142}:357-387

\bibitem{BGP} Bramon  A, Grau A, Pancheri G, Phys. Lett., 1995, {\bf 344}:240--244

\bibitem{BGP2} Bramon  A, Grau A, Pancheri G, Phys. Lett., 1995, {\bf 345}:263--268

\bibitem{taupaper} Benayoun  M, David P, DelBuono L, Leitner  O, O'Connell, H  B,
Eur. Phys. J., {\bf C55}:199--236

\bibitem{rad} Benayoun  M, DelBuono L, Eidelman S, Ivanchenko  V  N, O'Connell H  B,
Phys. Rev., 1999, {\bf D59}:114027

\bibitem{box} Benayoun  M, David P, DelBuono L, Leruste, P, O'Connell, H  B,
Eur. Phys. J., 2003, {\bf C31}:525--547

\bibitem{ExtMod1} Benayoun  M, David P, DelBuono L, Leitner  O,  arXiv:0907.4047
(hep--ph), 2009, accepted for publication in  Eur. Phys. J. C

\bibitem{ExtMod2} Benayoun  M, David P, DelBuono L, Leitner  O, arXiv:0907.5603
(hep--ph), 2009, submitted for publication in  Eur. Phys. J. C

\bibitem{BGPter} Bramon  A, Grau A, Pancheri G, Jose Luis Lucio M,
Phys. Lett., 2000, {\bf B486}:406--413

\bibitem{Marciano} Marciano W J, Sirlin A, Phys. Rev. Lett.,
1993, {\bf 71}:3629--3632

\bibitem{Cirigliano2} CiriglianoV, Ecker G,Neufeld H, Phys. Lett., 2001,
Phys. Lett., {\bf B513}:361-370

\bibitem{Cirigliano3} CiriglianoV, Ecker G,Neufeld H, JHEP,
2002, {\bf 08}:002

\bibitem{Barkov} Barkov  L  M  et al.,  Nucl. Phys., 1985, {\bf B256}:365--384

\bibitem{DM1} Quenzer  A  et al., Phys. Lett., 1978, {\bf B76}:512--516

\bibitem{CMD2-1995} Aulchenko V  M et al, Phys. Lett., 2002, {\bf B527}:161--172

\bibitem{CMD2-1995corr} Akhmetshin  R  R et al, Phys. Lett., 2004, {\bf B578}:285--289

\bibitem{CMD2-1998-1} Akhmetshin  R  R et al, Phys. Lett., 2007, {\bf B648}:28--38

\bibitem{CMD2-1998-2} Akhmetshin  R  R et al, JETP Lett., 2006, {\bf 84}:413--417

\bibitem{SND-1998} Achasov  M  N et al, J. Exp. Theor. Phys., 2006, {\bf 103}:380--384

\bibitem{CMD2Pg1999} Akhmetshin  R  R et al, Phys. Lett., 1999, {\bf B460}:242--247

\bibitem{CMD2Pg2005} Akhmetshin  R  R et al, Phys. Lett., 2005, {\bf B605}:23--36

\bibitem{SNDPg2000} Achasov  M  N et al, Eur. Phys. J., 2000, {\bf C12}:25--33

\bibitem{SNDPg2003} Achasov  M  N et al, Phys. Lett., 2003, {\bf B559}:171--178

\bibitem{SNDPg2007} Achasov  M  N et al, Phys. Rev., 2007, {\bf D76}:077101

\bibitem{CMD2-3p-1995}  Akhmetshin  R  R et al, Phys. Lett., 1995, {\bf B364}:199--206

\bibitem{CMD2-3p-1998}  Akhmetshin  R  R et al, Phys. Lett., 1995, {\bf B434}:426--436

\bibitem{CMD2-3p-2006}  Akhmetshin  R  R et al, Phys. Lett., 2006, {\bf B642}:203--209

\bibitem{SND3pionLow}  Achasov  M  N et al, Phys. Rev., 2003, {\bf D68}:052006

\bibitem{SND3pionHigh}  Achasov  M  N et al, Phys. Rev., 2002, {\bf D66}:032001

\bibitem{ND3pion-1991} Dolinsky  S  I et al, Phys. Rept., 1991, {\bf 202}:99--170

\bibitem{CMD3pion-1989} Barkov  L  M  et al,  BudkerINP preprint, 1989, 89-15, Novosibirsk

\bibitem{DM13pion-1979} Cordier A et al.,  Nucl. Phys., 1980, {\bf B172}:13

\bibitem{KLOE-2008} Ambrosino F et al.,  Phys. Lett., 2009, {\bf B670}:285--291

\bibitem{Cleo} Anderson  S  et al., Phys. Rev., 2000, {\bf D61}:112002

\bibitem{Aleph} Schael  S et al.,  Phys. Rept., 2005, {\bf 421}:191--284

\bibitem{Belle} Fujikawa  M et al., Phys. Rev., 2008,  {\bf D78}:072006

\bibitem{RPP2008} Amsler  C  et al., Phys. Lett., 2008, {\bf B667}:1--1340

\bibitem{Davier} Davier  M et al.,arXiv:0907.5603,(hep--ph), 2009, 
submitted for publication in  Eur. Phys. J. C

\bibitem{KLOE-2009}  Venanzoni  G  et al., arXiv:0906.4331, (hep--ex), 2009

\bibitem{Davier2} Davier  M et al., arXiv:0908.4300, (hep--ph), 2009

\end{thebibliography}
\end{document}